\documentstyle[12pt]{article}

\begin{document}
\setcounter{page}{1} \pagestyle{plain} \vspace{1cm}
\begin{center}
\Large{\bf UV/IR Mixing and Black Hole Thermodynamics}\\
\small \vspace{1cm}
{\bf Kourosh Nozari}$^{a,b}$\quad \quad and \quad \quad {\bf Behnaz Fazlpour$^{b}$ }\\
\vspace{0.5cm} $^{a}$ {\it Centre for Particle Theory, Durham
University, South Road, Durham DH1 3LE, UK}\\
$^{b}${\it Department of Physics, Faculty of Basic Sciences,
University of Mazandaran,\\
P. O. Box 47416-1467,
Babolsar, IRAN\\
e-mail: kourosh.nozari@durham.ac.uk}
\end{center}
\vspace{1.5cm}
\begin{abstract}
The goal of this paper is to investigate the final stage of black
hole evaporation process in the framework of Lorentz violating
Modified Dispersion Relations(MDRs). As a consequence of MDRs, the
high energy sector of the underlying field theory does not decouple
from the low energy sector, the phenomenon which is known as UV/IR
mixing. In the absence of exact supersymmetry, we derive a modified
dispersion relation which shows UV/IR mixing by a novel energy
dependence. Then we investigate the effects of these type of MDRs on
the thermodynamics of a radiating noncommutative Schwarzschild black
hole. The final stage of black hole evaporation obtained in this
framework is compared with existing pictures.
\end{abstract}

\newpage
\section{Introduction}
There are some evidences from alternative approaches to quantum
gravity which indicate that Lorentz symmetry is not an exact
symmetry of the nature[1-5]. One possible framework to incorporate
this Lorentz invariance violation in the equations of physics is the
modification of the standard dispersion relation[2,4,6-8]. The
resulting modified dispersion relations(MDRs) show some interesting
aspects of Planck scale physics such as UV/IR mixing. Based on this
idea, the high energy sector of the theory does not decouple from
the low energy sector. These features may reflect the fact that
spacetime at quantum gravity level has a granular structure. This
granular feature can describe the energy dependence of correction
terms to standard dispersion relation[7]. The effects of MDRs on
various aspects of quantum gravity problem have been studied
extensively[9-14]. From a phenomenological point of view, these MDRs
can be considered the basis of some important test theories which
can justify alternative approaches to quantum gravity
problem[15,16]. One of the important outcome of MDRs is the possible
interpretation of the astrophysical anomalies such as GZK
\footnote{The Greisen-Zatsepin-Kuzmin limit (GZK limit) is a
theoretical upper limit on the energy of cosmic rays from distant
sources[4,5,6,16,17,18]. } and $TeV$ photon anomaly[4,19]. Since
MDRs are common feature of all promising quantum gravity candidates,
it would be interesting to examine the effects of them on a key
problem of quantum gravity, that is, black hole thermodynamics.
Based on this view point, some authors have applied MDRs to the
formulation of black hole physics[9,10]. An elegant application of
MDRs to black hole physics and some related issues is provided by
Amelino-Camelia {\it et al}[9]. They have used a formulation of MDRs
which is common in existing literature and has the following form
\begin{equation}
 ({\vec p})^2\simeq {E}^2 - {\mu}^2 + {\alpha}_1 {L}_P
{E}^3 + {\alpha}_2 {L}_{P}^{2} E^4 + O({L}_P^3{E}^5)
\end{equation}
where $\mu$ is related to the rest mass and $\alpha_{i}$ are quantum
gravity model-dependent constants that may take different values for
different particles[7]. These type of MDRs have several implications
on the final stage of black hole evaporation. Comparison between the
results of MDRs for black hole thermodynamics and the standard
result of string theory, imposes some important constraints on the
functional form of MDRs as given in (1)[9,20]. For instance, as has
been shown in [10], only even powers of energy should be present in
relations such as (1).

Based on MDRs[9,10,20], the generalized uncertainty
principle[21-24], and also noncommutative geometry[25-27], the
current view point on the final stage of a black hole evaporation
can be summarized as follows:\\
{\it Black hole evaporates by emission of Hawking radiation in such
a way that in the final stage of evaporation, it reaches to a
maximum temperature before cooling down and finally reaches to a
stable remnant with zero entropy}.\\
The purpose of this paper is to re-examine black hole thermodynamics
within a combination of MDRs and space noncommutativity. Using a
general formulation of modified dispersion relations in the language
of noncommutative geometry, we obtain a MDR which contains a novel
energy dependence relative to relation (1). This type of MDR
consists of modification terms which are functions of inverse of
powers of energy and show an explicit UV/IR mixing. We apply our MDR
to the
issue of black hole thermodynamics and compare our results with existing picture.\\
The paper is organized as follows: In section $2$ we use
noncommutative space framework to obtain a new MDR with a novel
energy dependence. Section $3$ applies our MDR to the issue of black
hole thermodynamics. The paper follows by discussion and results in
section $4$.

\section{Modified Dispersion Relations and UV/IR Mixing}
In this section, using the notion of space noncommutativity, we find
a new modified dispersion relation indicating explicit UV/IR mixing.
A noncommutative space can be defined by the coordinate operators
satisfying the following commutation relation[28-32]
\begin{equation}
[x_\mu ,x_\nu]=i\theta_{\mu \nu}+i\rho_{\mu \nu}^{\beta}x_{\beta}
\end{equation}
In the spacial case where $\rho_{\mu \nu}^{\beta}$ vanishes, we find
the canonical noncommutative spacetime with the following algebraic
structure
\begin{equation}
[x_\mu ,x_\nu]=i\theta_{\mu \nu}
\end{equation}
where $\hat x$'s are the coordinate operators and $ \theta_{\mu
\nu}$ is an antisymmetric matrix whose elements
have dimension of $(length)^2$.\\
Modification of standard field theory due to spacetime
noncommutativity has been studied extensively(see for example
[30,32,33]). One important consequence of noncommutative field
theory is emergence of the so called UV/IR mixing; the high energy
sector of the theory does not decouple from the low energy sector.
This UV/IR mixing can be addressed via modification of standard
dispersion relations. Within the canonical noncommutative field
theory, this modified dispersion relation attains the following
form[7]
\begin{equation}
m^2\simeq E^2-\vec{p} ^{2}+\frac{\alpha_{1}}{p^{\mu }\theta_{\mu
\nu}\theta^{\nu \sigma}p_{\sigma} } +\alpha_2 m^2 \ln(p^{\mu
}\theta_{\mu \nu}\theta^{\nu \sigma}p_{\sigma})+...,
\end{equation}
where $\theta_{ij} = \frac{1}{2} \epsilon_{ijk} \theta^k$. The
$\alpha_{i}$ are parameters dependent on various aspects of the
field theory and can take different values for different particles
since dispersion relation is not universal. Using the identity
\begin{equation}
\epsilon_{ijr} \epsilon_{iks} = \delta_{jk} \delta_{rs} -
\delta_{js} \delta_{rk},
\end{equation}
one finds
\begin{equation}
p^{\mu }\theta_{\mu \nu}\theta^{\nu
\sigma}p_{\sigma}=-\frac{1}{4}\Big(p^2
\theta^2-(\vec{p}.\vec{\theta})^2\Big).
\end{equation}
Therefore, equation (4) can be written as follows
\begin{equation}
E^2\simeq m^2+\vec{p} ^{2}+\frac{4 \alpha_{1}}{\Big(p^2
\theta^2-(\vec{p}.\vec{\theta})^2\Big)}-\alpha_2 m^2 \ln
\Big[-\frac{1}{4}\Big(p^2 \theta^2-(\vec{p}.\vec{\theta})^2
\Big)\Big].
\end{equation}
This MDR can be singular in the infrared regime as a result of
UV/IR mixing. In the case of exact supersymmetry, part of this
infrared singularity can be removed by setting $\alpha_{1}=0$. The
case with $\alpha_{1}\neq 0$ has not been considered in literature
but as we will show it has some novel implication in the spirit of
black hole thermodynamics. So, in which follows we consider the
non-supersymmetric case where $\alpha_{1}\neq 0$. On the other
hand, except for situation where $\vec{p}.\vec{\theta}=p\theta$,
the quantity $-\frac{1}{4}\Big(p^2
\theta^2-(\vec{p}.\vec{\theta})^2\Big)$ is negative, therefore the
last term is imaginary. As an example of this situation, note that
phonons in a fluid flow can propagate with an MDR which shows
imaginary terms if viscosity is taken into account. In this paper
we don't consider these extreme situations therefore in which
follows, we consider the case where $\alpha_{2}=0$. The parameter
$m$ is directly related to the rest energy, and in the high energy
regime we can neglect it. Considering these points, we find
\begin{equation}
E^2\simeq \vec{p} ^{2}+\frac{4 \alpha_{1}}{\Big(p^2
\theta^2-(\vec{p}.\vec{\theta})^2\Big)}
\end{equation}
where $p^2={\vec p}.{\vec p}\,$ and $\,\theta^2={\vec \theta}.{\vec
\theta}$ . If we set $\theta_3=\theta$ and assuming that remaining
components of $\theta$ all vanish (which can be done by a rotation
or a re-definition of the coordinates), then ${\vec p}.{\vec
\theta}=p_z \theta$. In this situation equation (8) can be written
as follows
\begin{equation}
E^2\simeq (p_x^2+p_y^2+p_z^2)+\frac{4 \alpha_{1}}{(p_x^2+p_y^2)
\theta^2}.
\end{equation}
Assuming an isotropic case where $p_x=p_y=p_z=\tilde{p}$, we find
\begin{equation}
3\tilde{p}^2+\frac{2 \alpha_{1}}{\tilde{p}^2 \theta^2}-E^2=0.
\end{equation}
This equation has two solutions for $\tilde{p}^2$. Only one of these
solutions is acceptable since in standard limit where $\alpha_1=0$
we should recover $3\tilde{p}^2=E^2$ (note that we have omitted the
rest mass from our calculations). This solution is
\begin{equation}
\tilde{p}^2=\frac{1}{6}\Big(E^2+\sqrt{E^4-\frac{24
\alpha_1}{\theta^2}} \Big)
\end{equation}
or
\begin{equation}
p^2=\frac{1}{2}\Big(E^2+\sqrt{E^4-\frac{24 \alpha_1}{\theta^2}}
\Big).
\end{equation}
We expand this relation up to second order of $\alpha_1$ to find
\begin{equation}
p^2=E^2-\frac{6 \alpha_1}{\theta^2 E^2}-\frac{36
\alpha_1^2}{\theta^4 E^6}+{\cal{O}}(\frac{ \alpha_1^3}{\theta^6
E^{10}})
\end{equation}
Our forthcoming arguments are based on this result. It provides an
energy dependence which has not been pointed out in existing
literature explicitly. Note that based on Heisenberg uncertainty
principle, since always $E\geq \frac{1}{\delta x}$  where $\delta x$
is particle position uncertainty, relation (13) is well defined. In
the existing literature, following analysis of Amelino-Camelia {\it
et al} [9], in most cases, the authors have led to consider a
dispersion relation of the type
\begin{equation}
\vec{p}^2\simeq E^2+\beta_1 L_p E^3+\beta_2 L_p^2
E^4+{\cal{O}}(L_p^3 E^5)
\end{equation}
where the coefficients $\beta_i$ can take different values in
different quantum gravity proposals. In this type of MDRs there is
no dependence to the inverse powers of $E$. However, our result
given by (13) contains modification terms with powers of inverse of
$E$. Note also that our MDR contains even powers of energy which is
in agreement with our previous finding[10]. We proceed in the line
of Amelino-Camelia {\it et al} approach[9] but instead of their MDR
as given by (14), we use our MDR (13) which mainly considers the IR
limit of the noncommutative field theory. In this manner, some novel
results are obtained which illuminate further the final stage of
black hole evaporation. Due to different energy dependence of our
MDR, we expect that thermodynamics of black hole obtained within
this framework differs from black hole thermodynamics obtained in
other MDRs framework. For instance, we will observe that in this
framework there is no logarithmic correction to Bekenstein-Hawking
entropy-area relation. This difference may reflect some aspects of
UV/IR mixing and related quantum gravity phenomena.

The relation between $dp$ and $dE$ obtained from (13) is as follows
\begin{equation}
dp=\Big[1+a_1 \Big(\frac{\alpha_1}{\theta^2 E^4}\Big)+a_2
\Big(\frac{\alpha_1}{\theta^2 E^4}\Big)^2+a_3
\Big(\frac{\alpha_1}{\theta^2 E^4}\Big)^3 \Big]dE
\end{equation}
where the coefficients $a_i$ are constant and we consider terms only
up to third order of $\alpha_1$.\\
According to Heisenberg's uncertainty principle, in order to measure
the particle position with precision $\delta x$ one should use a
photon with momentum uncertainty $\delta p\geq \frac{1}{\delta x}$.
Within our framework and considering this point, one is led to the
following relation
\begin{equation}
E \geq \frac{1}{\delta x}\Bigg[1-a_1 \Big(\frac{\alpha_1 (\delta
x)^4}{\theta^2}\Big)+(a_1^2-a_2) \Big(\frac{\alpha_1 (\delta
x)^4}{\theta^2}\Big)^2+(2 a_1 a_2 -a_3 -a_1^3) \Big(\frac{\alpha_1
(\delta x)^4}{\theta^2}\Big)^3\Bigg].
\end{equation}
For the standard case where $a_{i}=0$, this equation simplifies to
$E\geq\frac{1}{\delta x}$. In which follows, we show that the
modification of dispersion relation of the type (13), can lead to
corrections of standard relations for entropy and temperature of the
black hole, i. e. standard Bekenstein-Hawking entropy-area relations
$S=\frac{A}{4}$ and $T=\frac{1}{8 \pi M}$ will be modified as a
result of UV/IR mixing.

\section{Black Hole Thermodynamics with Modified Dispersion Relation}
The Bekenstein argument suggests that the entropy of a black hole
should be proportional to its area of event horizon. The minimum
increase of area when the black hole absorbs a classical particle of
energy $E$ and size $s$ is $\Delta A\geq 8 \pi E s$\,\,[9]. When
black hole absorbs a quantum particle of size $s$, uncertainty in
position of the particle will be $\delta x$ where $s\sim \delta x$.
Considering a calibration factor as $\frac{\ln 2}{2 \pi}$, we find
\begin{equation}
\Delta A\geq 4(\ln 2)E \delta x;
\end{equation}
In the standard case $E\sim \frac{1}{\delta x}$, which leads to
\begin{equation}
\Delta A\geq 4(\ln 2).
\end{equation}
Using the fact that the minimum increase of entropy is $\ln 2$\,(one
bit of information) and it is independent of the area $A$, one find
\begin{equation}
\frac{dS}{dA}\simeq \frac{(\Delta S)_{min}}{(\Delta
A)_{min}}\simeq\frac{1}{4}.
\end{equation}
Integration leads to standard Bekenstein result
\begin{equation}
S\simeq\frac{A}{4}.
\end{equation}
To calculate entropy in the presence of UV/IR mixing, we use
relations (15) and (16) to find
\begin{equation}
\Delta A\geq 4 (\ln 2)\Bigg[1-a_1 \Big(\frac{\alpha_1 (\delta
x)^4}{\theta^2}\Big)+(a_1^2-a_2) \Big(\frac{\alpha_1 (\delta
x)^4}{\theta^2}\Big)^2+(2 a_1 a_2 -a_3 -a_1^3) \Big(\frac{\alpha_1
(\delta x)^4}{\theta^2}\Big)^3\Bigg].
\end{equation}
Now to calculate entropy and temperature of black hole we need the
radius of event horizon $r_H$. There are two relatively different
approaches to find noncommutative radius of the event
horizon[25,27]. These two approaches are based on two different
view points to make general relativity noncommutative(see [25] and
[34] for further details). Here we use Nicollini {\it et al }
approach to find noncommutative radius of event horizon[25]. It
has been shown that noncommutativity eliminates point-like
structures in favor of smeared objects in flat spacetime. As
Nicolini {\it et al} have shown, the effect of smearing is
mathematically implemented as a substitution rule: position
Dirac-delta function is replaced everywhere with a Gaussian
distribution of minimal width $\sqrt{\theta}$. In this framework,
they have chosen the mass density of a static, spherically
symmetric, smeared, particle-like gravitational source as follows
\begin{equation}
\rho_\theta(r)=\frac{M}{(2\pi\theta)^{\frac{3}{2}}}\exp(-\frac{r^2}{4\theta})
\end{equation}
As they have indicated, the particle mass $M$, instead of being
perfectly localized at a point, is diffused throughout a region of
linear size $\sqrt{\theta}$. This is due to the intrinsic
uncertainty as has been shown in the coordinate commutators (3).
This kind of matter source results the following static, spherically
symmetric, asymptotically Schwarzschild solution of the Einstein
equations[25,26]
\begin{equation}
ds^2=\Bigg(1-\frac{2M}{r\sqrt{\pi}}\gamma\Big(\frac{1}{2},\frac{r^2}{4\theta}\Big)\Bigg)dt^2-
\Bigg(1-\frac{2M}{r\sqrt{\pi}}\gamma\Big(\frac{1}{2},\frac{r^2}{4\theta}\Big)\Bigg)^{-1}dr^2-r^2
(d\vartheta^2+sin^2\vartheta d\phi^2)
\end{equation}
where $\gamma\Big(\frac{1}{2},\frac{r^2}{4\theta}\Big)$ is the lower
incomplete Gamma function:
\begin{equation}
\gamma\Big(\frac{1}{2},\frac{r^2}{4\theta}\Big)\equiv\int_0^{\frac{r^2}{4\theta}}t^{\frac{1}{2}}e^{-t}dt
\end{equation}
The event horizon of this metric can be found where $g_{00} ( r_H )
= 0$,
\begin{equation}
r_H=\frac{2M}{\sqrt{\pi}}\gamma\Big(\frac{1}{2},\frac{r_H^2}{4\theta}\Big)
\end{equation}
As it is obvious from this equation, the effect of noncommutativity
in the large radius regime can be neglected, while at short distance
regime one expects significant changes due to the spacetime
fuzziness.

After determination of noncommutative radius of black hole event
horizon, we return to our original argument on black hole
thermodynamics. When a particle falls into the black hole event
horizon, the particle position uncertainty will be $\delta x\sim
r_H$, where $r_H$ is the Schwarzchild radius in noncommutative
spacetime. Therefore we have from (21)
\begin{equation}
\Delta A\geq 4 (\ln 2)\Bigg[1-a_1 \Big(\frac{\alpha_1
}{\theta^2}\Big)r_H^{4} +(a_1^2-a_2) \Big(\frac{\alpha_1
}{\theta^2}\Big)^2r_H^{8}+(2 a_1 a_2 -a_3 -a_1^3)
\Big(\frac{\alpha_1 }{\theta^2}\Big)^3r_H^{12}\Bigg]
\end{equation}
defining $A=4 \pi r_H^2$, equation (26) takes the following form
$$\Delta A\geq 4 (\ln 2)\Bigg[1-a_1 \Big(\frac{\alpha_1
}{\theta^2}\Big)\Big(\frac{A
\gamma^2(\frac{1}{2},\frac{r_H^2}{4\theta})}{4
\pi^2}\Big)^2+(a_1^2-a_2) \Big(\frac{\alpha_1
}{\theta^2}\Big)^2\Big(\frac{A
\gamma^2(\frac{1}{2},\frac{r_H^2}{4\theta})}{4 \pi^2}\Big)^4$$
\begin{equation}
+(2 a_1 a_2 -a_3 -a_1^3) \Big(\frac{\alpha_1
}{\theta^2}\Big)^3\Big(\frac{A
\gamma^2(\frac{1}{2},\frac{r_H^2}{4\theta})}{4
\pi^2}\Big)^6\Bigg].
\end{equation}
Now the entropy of black hole can be calculated as follows
$$\frac{dS}{dA}\simeq \frac{(\Delta S)_{min}}{(\Delta
A)_{min}}$$
\begin{equation}
\simeq \frac{1}{4}\Bigg[1+a_1 \Big(\frac{\alpha_1
}{\theta^2}\Big)\Big(\frac{A
\gamma^2(\frac{1}{2},\frac{r_H^2}{4\theta})}{4 \pi^2}\Big)^2+a_2
\Big(\frac{\alpha_1 }{\theta^2}\Big)^2\Big(\frac{A
\gamma^2(\frac{1}{2},\frac{r_H^2}{4\theta})}{4 \pi^2}\Big)^4+ a_{3}
\Big(\frac{\alpha_1 }{\theta^2}\Big)^3\Big(\frac{A
\gamma^2(\frac{1}{2},\frac{r_H^2}{4\theta})}{4 \pi^2}\Big)^6\Bigg].
\end{equation}
In terms of event horizon radius, this relation can be written as
\begin{equation}
\frac{dS}{dr_H}\simeq \Bigg[2\pi r_H+ \frac{2 a_1}{\pi}
\Big(\frac{\sqrt{\alpha_1}
\gamma^2(\frac{1}{2},\frac{r_H^2}{4\theta})}{\theta}\Big)^2
r_H^5+\frac{2 a_2}{\pi^3} \Big(\frac{\sqrt{\alpha_1}
\gamma^2(\frac{1}{2},\frac{r_H^2}{4\theta})}{\theta}\Big)^4
r_H^9+\frac{2a_3}{\pi^5} \Big(\frac{\sqrt{\alpha_1}
\gamma^2(\frac{1}{2},\frac{r_H^2}{4\theta})}{\theta}\Big)^6
r_H^{13}\Bigg],
\end{equation}
which integration gives
$$S\simeq \frac{A}{4}+\int\Bigg[\frac{2 a_1}{\pi}
\Big(\frac{\sqrt{\alpha_1}
\gamma^2(\frac{1}{2},\frac{r_H^2}{4\theta})}{\theta}\Big)^2
r_H^5+\frac{2 a_2}{\pi^3} \Big(\frac{\sqrt{\alpha_1}
\gamma^2(\frac{1}{2},\frac{r_H^2}{4\theta})}{\theta}\Big)^4
r_H^{9}$$
\begin{equation}
+\frac{2a_3}{\pi^5} \Big(\frac{\sqrt{\alpha_1}
\gamma^2(\frac{1}{2},\frac{r_H^2}{4\theta})}{\theta}\Big)^6
r_H^{13}\Bigg]dr_H.
\end{equation}
The first term in the right hand side is the standard Bekenstein
entropy. But, integration of second term has no closed form and can
be calculated numerically. This result shows that black hole entropy
in the presence of MDR(UV/IR mixing) has a complicated form and this
form is different from the entropy-area relation obtained in other
quantum gravity-based approaches(see for example results of
Amelino-Camelia {\it et al} in [9]). Figure $1$ shows the numerical
calculation of entropy-event horizon relation for an evaporating
black hole in Bekenstein-Hawking and the noncommutative geometry
view points\footnote{As we have mentioned in page 4, $\alpha_{i}$
coefficients are parameters dependent on various aspects of the
field theory and can take different values for different particles
since dispersion relation is not universal. To have a qualitative
behavior of our thermodynamical quantities, we have set
$\alpha_{1}=1$ in figures. On the other hand, since we have set
$\hbar=1$, our noncommutative commutation relations are given by
(3). With $\hbar \neq 1$ we will find $[x_\mu ,x_\nu]=i\hbar
\theta_{\mu \nu}$ where in this case $\theta_{\mu\nu}$ will have
dimension of $\frac{(length)^2}{\hbar}$. In this regard, we have set
$\theta=1$ in our numerical calculations. This is a common approach
to find qualitative description of black hole thermodynamical
quantities. Currently no concrete values of these quantities are in
hand since these parameters are quantum gravity-model dependent.}.
As this figure shows, within noncommutative geometry approach, black
hole in its final stage of evaporation reaches to a zero entropy
remnant.

\begin{figure}[htp]
\begin{center}
\includegraphics{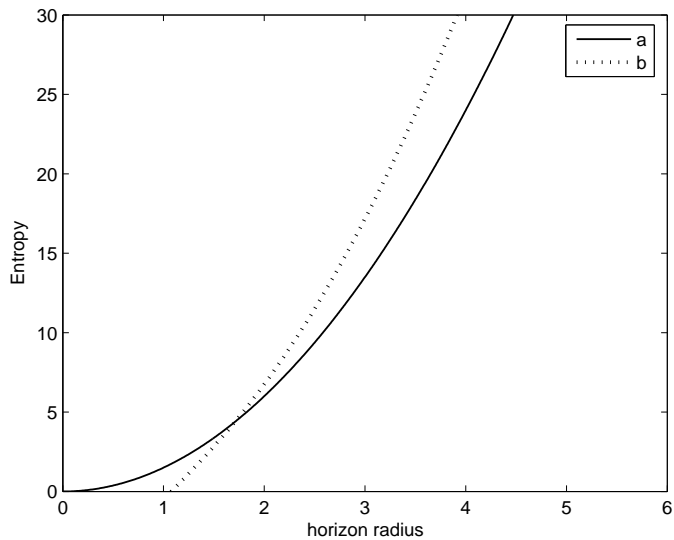}
\end{center}
\vspace{7 cm}
 \caption{\small {Black hole entropy versus its radius of event horizon:
  a)standard Bekenstein-Hawking result, and  b)noncommutative space result.
  We have set $\theta=1$ and $\alpha_{1}=1$. All $a_{i}$ coefficients
  are assumed to be positive. }}
 \label{fig:1}
\end{figure}

This result is in agreement with existing literature. However in our
case we reach a remnant with larger mass relative to existing
results such as result of Nicolini {\it et al}[25]. To proceed
further, we discuss two extreme limits: commutative limit and highly
noncommutative limit.
\begin{itemize}
\item
In the large radius regime $\frac{r_H^2}{4\theta}\gg 1$\,\,
(commutative regime), the $\gamma$ function can be solved to the
first order of approximation to find
\begin{equation}
\gamma(\frac{1}{2},\frac{r_H^2}{4\theta})=-\frac{2\sqrt{\theta}}{r_H}e^{-\frac{r_H^2}{4\theta}}.
\end{equation}
Therefore from equation (29) we find
\begin{equation}
\frac{dS}{dr_H}\simeq 2\pi r_H\Bigg[1+ \frac{ a_1}{\pi^2} \Big(16
\alpha_1 e^{-\frac{r_H^2}{\theta}}\Big)+ \frac{ a_2}{\pi^4} \Big(16
\alpha_1 e^{-\frac{r_H^2}{\theta}}\Big)^2+ \frac{ a_3}{\pi^6}
\Big(16 \alpha_1 e^{-\frac{r_H^2}{\theta}}\Big)^3 \Bigg].
\end{equation}
In this case entropy can be calculated as follows
\begin{equation}
S\simeq \frac{A}{4}- \frac{ a_1 \theta}{\pi} \Big(16 \alpha_1
e^{-\frac{A}{4\pi\theta}}\Big)- \frac{ a_2 \theta}{2 \pi^3} \Big(16
\alpha_1 e^{-\frac{A}{4 \pi\theta}}\Big)^2- \frac{
a_3\theta}{3\pi^5} \Big(16 \alpha_1 e^{-\frac{A}{4\pi
\theta}}\Big)^3 .
\end{equation}
We see from this relation that for large radii with respect to
$\sqrt{\theta}$, the effect of noncommutativity of spacetime
is exponentially small.\\
\item
In the opposite limit where $r_H\simeq\sqrt{\theta}$, the structure
of spacetime is fuzzy and the $\gamma$ function in this limit
(noncommutative limit) can be solved as
\begin{equation}
\gamma(\frac{1}{2},\frac{r_H^2}{4\theta})=\frac{r_H}{\sqrt{\theta}}e^{-\frac{r_H^2}{4\theta}},
\end{equation}
In this case equation (29) leads to the following expression
\begin{equation} \frac{dS}{dr_H}\simeq 2\pi r_H\Bigg[1+ \frac{
a_1}{\pi^2} \Big( \frac{\alpha_1 r_H^8
e^{-\frac{r_H^2}{\theta}}}{\theta^4} \Big)+ \frac{ a_2}{\pi^4} \Big(
\frac{\alpha_1 r_H^8e^{-\frac{r_H^2}{\theta}}}{\theta^4} \Big)^2+
\frac{a_3}{\pi^6} \Big( \frac{\alpha_1
r_H^8e^{-\frac{r_H^2}{\theta}}}{\theta^4} \Big)^3 \Bigg],
\end{equation}
and integration leads to the following result for black hole entropy
in strong noncommutative limit
$$S\simeq \frac{A}{4}-\frac{a_1 \alpha_1}{\pi}\Bigg[
24\theta+24\Big(\frac{A}{4\pi}\Big)+\frac{12}{\theta}\Big(\frac{A}{4\pi}\Big)^2
+\frac{4}{\theta^2}\Big(\frac{A}{4\pi}\Big)^3+\frac{1}{\theta^3}
\Big(\frac{A}{4\pi}\Big)^4\Bigg]e^{-\frac{A}{4\pi\theta}}$$
$$-\frac{\mu a_2 \alpha_1^2}{\pi^3}\Bigg[
\frac{3}{2}\theta+3\Big(\frac{A}{4\pi}\Big)+\frac{3}{\theta}\Big(\frac{A}{4\pi}\Big)^2
+\frac{2}{\theta^2}\Big(\frac{A}{4\pi}\Big)^3+\frac{1}{\theta^3}
\Big(\frac{A}{4\pi}\Big)^4\Bigg]e^{-\frac{2A}{4\pi\theta}}$$
\begin{equation}
-\frac{\nu a_3 \alpha_1^3}{\pi^5}\Bigg[
\frac{8}{27}\theta+\frac{8}{9}\Big(\frac{A}{4\pi}\Big)+\frac{4}{3\theta}
\Big(\frac{A}{4\pi}\Big)^2
+\frac{4}{3\theta^2}\Big(\frac{A}{4\pi}\Big)^3+\frac{1}{2\theta^3}
\Big(\frac{A}{4\pi}\Big)^4\Bigg]e^{-\frac{3A}{4\pi\theta}},
\end{equation}
where we have considered only terms up to fourth order of $A$ in
brackets and $\mu$ and $\nu$ are constant. This is an interesting
result which shows the modified entropy of black hole in
noncommutative geometry. Two main characteristic feature of this
relation are: it has Bekenstein-Hawking result as commutative limit
but it contains no logarithmic correction term. It is commonly
believed that entropy should have a series expansion as follows(see
[10] and references therein)
\begin{equation}
S\simeq \frac{A}{4}-\frac{\pi\zeta}{2} \ln{\frac{
A}{4}}+\sum_{n=1}^{\infty}c_{n}\Big(\frac{4}{A}\Big)^{n}+\cal{C}.
\end{equation}
Our analysis shows that in the presence of UV/IR mixing there are
some deviation from this standard result since there is no
logarithmic correction term and also there are power of event
horizon area instead of inverse of power of event horizon area. The
presence of logarithmic correction term which is a matter of debate
in the literatures now finds a natural solution in MDRs framework.
\end{itemize}

Using the first law of black hole thermodynamics $dS=\frac{dM}{T}$,
we can calculate the corrected temperature in the framework of our
analysis. Considering $\frac{dS}{dM}=\frac{1}{T}$ and $r_H=2M$, the
corrected temperature in noncommutative spacetime is obtained as
follows
$$T\simeq \frac{1}{8\pi M}\Bigg[1-a_1 \Big(\frac{\sqrt{\alpha_1}
\gamma^2(\frac{1}{2},\frac{r_H^2}{4\theta})}{\pi\theta}\Big)^2(4
M^2)^2+(a_1^2-a_2) \Big(\frac{\sqrt{\alpha_1}
\gamma^2(\frac{1}{2},\frac{r_H^2}{4\theta})}{\pi\theta}\Big)^4 (4
M^2)^4$$
\begin{equation}
+(2a_1 a_2- a_1^3-a_3) \Big(\frac{\sqrt{\alpha_1}
\gamma^2(\frac{1}{2},\frac{r_H^2}{4\theta})}{\pi\theta}\Big)^6(4
M^2)^6\Bigg].
\end{equation}
For semiclassical limit $\frac{r_H^2}{4\theta}\gg 1$, the
temperature using equation (31) takes the following form
\begin{equation}
T_H \simeq \frac{1}{4\pi r_H}\Bigg[1- a_1 \Big(\frac{16
\alpha_1}{\pi^2} e^{-\frac{r_H^2}{\theta}}\Big)+ (a_1^2-a_2)
\Big(\frac{16 \alpha_1}{\pi^2}
e^{-\frac{r_H^2}{\theta}}\Big)^2+(2a_1 a_2+2a_1^2-a_3)
\Big(\frac{16 \alpha_1}{\pi^2} e^{-\frac{r_H^2}{\theta}}\Big)^3
\Bigg],
\end{equation}
which leads to the standard relation for Hawking temperature
\begin{equation}
T_H = \frac{1}{4\pi r_H}.
\end{equation}
In noncommutative limit where $r_H\simeq \sqrt{\theta}$, using
equation (34) the temperature can be written as follows
\begin{equation}
T_H\simeq \frac{1}{4\pi r_H}\Bigg[1-a_1 \Big( \frac{\alpha_1 r_H^8
e^{-\frac{r_H^2}{\theta}}}{\pi^2\theta^4} \Big)+ (a_1^2-a_2) \Big(
\frac{\alpha_1 r_H^8e^{-\frac{r_H^2}{\theta}}}{\pi^2\theta^4}
\Big)^2+ (2a_1 a_2- a_1^3-a_3) \Big( \frac{\alpha_1
r_H^8e^{-\frac{r_H^2}{\theta}}}{\pi^2\theta^4} \Big)^3 \Bigg],
\end{equation}
which has very different form from existing picture but as we will
see its general behavior with respect to event horizon radius is the
same as existing picture. In the commutative limit, when the horizon
radius decreases, the temperature increases and diverges. In the
opposite limit, when we consider fuzzy spacetime, the black hole
before cooling down to absolute zero, reaches to a finite maximum
temperature. Figure 2 shows the temperature-event horizon relation
for an evaporating black hole in Bekenstein-Hawking and the
noncommutative geometry view points. This result is in agreement
with the results of Nicolini {\it et al} which have computed black
hole temperature in a different view point[25].

\begin{figure}[htp]
\begin{center}
\includegraphics{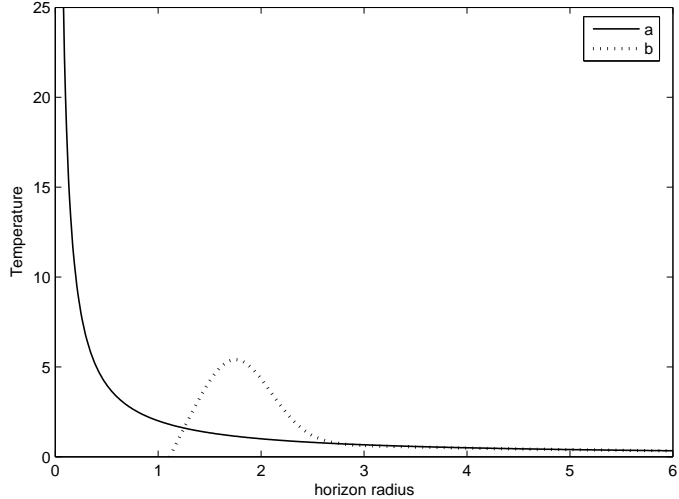}
\end{center}
\vspace{6 cm}
 \caption{\small {Black hole temperature versus its radius of event horizon in
 two different regimes:  a)standard Bekenstein-Hawking result, and
 b)noncommutative space result. We have set $\theta=1$ and $\alpha_{1}=1$. All $a_{i}$
 coefficients are assumed to be negative.}}
 \label{fig:2}
\end{figure}

\begin{figure}[htp]
\begin{center}
\includegraphics{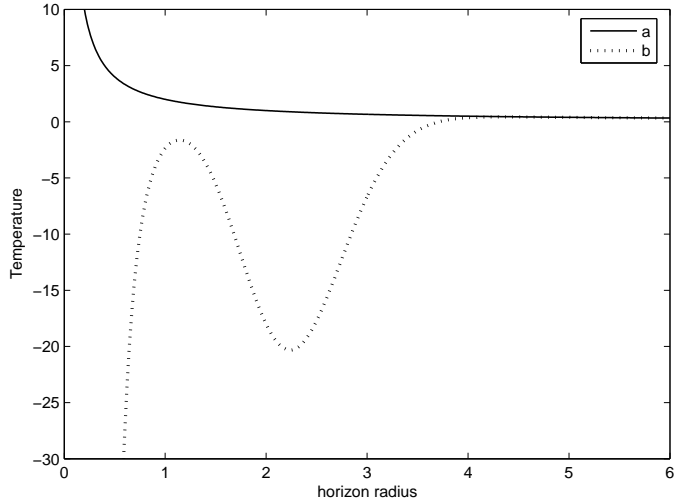}
\end{center}
\vspace{5 cm}
 \caption{\small {Black hole temperature versus its radius of event horizon in two different
 regimes and for the case of possible negative temperature:
 a)standard Bekenstein-Hawking result, and
 b)noncommutative space result. We have set $\theta=1$ and $\alpha_{1}=1$ but all
 coefficients $a_{i}$ are assumed to be positive.}}
 \label{fig:2}
\end{figure}

One point should be stressed here: for some values of constants in
relation (41), it is possible to have negative temperature for final
state of black hole evaporation(see figure 3). It is a well-known
issue in condensed matter physics that under certain conditions, a
closed system can be described by a negative temperature, and,
surprisingly, be hotter than the same system at any positive
temperature. Recently Park has shown that the Hawking temperature of
exotic black holes and the black holes in the three-dimensional
higher curvature gravities can be negative[35].In our opinion,
possible negative temperature of black holes in their final stage of
evaporation is a signature of anti-gravitation or may reflect the
fact that current extensive thermodynamics is not sufficient to
describe this extreme situation. Non-extensive thermodynamics of
Tsallis[36] may provide a better framework for this extra-ordinary
situation.

\section{Summary and Conclusion}
The UV/IR mixing is an outcome of spacetime noncommutativity. In a
noncommutative quantum field theory, the high energy sector of the
theory does not decouple from the low energy sector. This is the
main modification of standard field theory in noncommutative spaces.
This UV/IR mixing can be addressed in modification of standard
dispersion relation. The very important outcome of these modified
dispersion relations is the obvious Lorentz invariance violation at
high energies. We have tried to incorporate these important notions
to the issue of black hole evaporation. In the absence of
supersymmetry, we have found a modified dispersion relation which
has a novel energy dependence and highlights IR sector of the
theory. Then using the standard argument of Bekenstein, we have
calculated entropy and temperature of $TeV$ black holes using our
new MDR. We have considered two extreme limits and in each case
entropy and temperature of black hole are calculated as a function
of event horizon radius. In the course of calculation, the main
problem was the choice of noncommutative radius of event horizon. We
have solved this problem by replacing position Dirac delta function
by a smeared Gaussian distribution as has been pointed in [25]. The
overall behavior of our numerical solutions are the same of existing
results but there are some important differences. There is no
logarithmic correction term in our entropy-area relation. Also, we
find a remnant with larger mass relative to existing results.

The temperature behavior shows that noncommutativity plays the same
role in general relativity as in quantum field theory, i. e.,
removes short distance divergences. Note also that Hawking radiation
back-reaction should be considered to explain reduction of
temperature in final stage of evaporation. In commutative case one
expects relevant back-reaction effects during the terminal stage of
evaporation because of huge increase of temperature. In our
noncommutative case, the role of noncommutativity is to cool down
the black hole in final stage. As a consequence, there is a
suppression of quantum back reaction since the black hole emits less
and less energy. Eventually, back-reaction my be important during
the maximum temperature phase. Note also that, as a common belief in
existing literature, the final point of evaporation is a
zero-temperature $TeV$ remnants as figure $2$ shows. This is a
direct consequence of minimal length due to additional gravitational
uncertainty[21] and also noncommutative geometry which gives a
granular structure to the spacetime manifold[38,39].

In some special circumstances our relation for temperature gives a
negative temperature for a period of final stage of black hole
evaporation. This negative temperature can be explained as follows:
generally, negative absolute temperatures can be achieved by
crossing very high temperatures. It may be a signature of
anti-gravitation(repulsive behavior) or white-hole as recently has
been pointed by Castro[37]. In this framework, negative temperatures
( but positive entropy ) are inherently associated with the
repulsive gravity white-hole picture. Castro has argued that these
extra-ordinary problem of having negative temperatures can be
resolved by shifting of horizon location. Physically, these negative
temperature may be inherent in the failure of standard
thermodynamics in this extreme situation[34]. Non-extensive
thermodynamics may provide a better framework for this situation.
Therefore, although the general behaviors of our solutions are the
same as existing picture, they can motivate some new issues in the
spirit of black hole thermodynamics.\\

{\bf Acknowledgement}\\
The main part of this work has been done during KN sabbatical leave
at Durham University, UK. He would like to appreciate members of the
Centre for Particle Theory at Durham University, specially Professor
Ruth Gregory, for hospitality. We would also to appreciate referee
for his/her important contribution to this work.

\end{document}